%% file: main.tex
\documentclass[runningheads]{llncs}
\usepackage{graphicx}
\usepackage{pgf,tikz}
\usepackage{enumitem}
\usepackage{todonotes}
\usepackage{booktabs}
\newcommand{\mypar}[1]{\smallskip\vspace{0.3em}\noindent\textbf{#1.}}
\begin{document}
\usetikzlibrary{arrows,shapes.geometric,positioning,matrix}

\hyphenation{dash-boards}
\title{Towards a Theory on \\ Process Automation Effects}
%
%
\author{Hoang Vu\inst{1,3} \and
Jennifer Haase\inst{4,5} \and
Henrik Leopold\inst{2,3} \and
Jan Mendling\inst{4,5,6}\thanks{The research by Jennifer Haase and Jan Mendling was supported by the Einstein Foundation Berlin under grant EPP-2019-524 and by the German Federal Ministry of Education and Research under grant 16DII133.}%
\authorrunning{Vu et al.}
%
\institute{SAP, Walldorf, Germany \\ \email{h.vu@sap.com} \and
Kühne Logistics University, Hamburg, Germany \\
\email{henrik.leopold@the-klu.org} \and
Hasso Plattner Institute, University of Potsdam, Germany\\
\email{} \and
Humboldt-Universität zu Berlin, Berlin, Germany \\\email{\{jennifer.haase|jan.mendling\}@hu-berlin.de} \and
Weizenbaum Institute, Berlin, Germany \and
WU Vienna, Vienna, Austria \\
}}
\maketitle              
\begin{abstract}
Process automation is a crucial strategy for improving business processes, but little attention has been paid to the effects that automation has once it is operational. This paper addresses this research problem by reviewing the literature on human-automation interaction. Although many of the studies in this field have been conducted in different domains, they provide a foundation for developing propositions about process automation effects. Our analysis focuses on how humans perceive automation technology when working within a process, allowing us to propose an effective engagement model between technology, process participants, process managers, and software developers. This paper offers insights and recommendations that can help organizations optimize their use of process automation. We further derive novel research questions for a discourse within the process automation community.

\end{abstract}
\section{Introduction}
\input{sections/intro}

\section{Background}
\input{sections/background}

\section{Research Method}
\input{sections/methodology}

\section{Findings}
\input{sections/findings}

\section{Discussion}
\input{sections/discussion}

\section{Conclusion}
\input{sections/conclusion}


\end{document}

%% file: sections/intro.tex
Business process management (BPM) is concerned with the continuous improvement of business processes~\cite{dumas2018fundamentals}. Process improvements can be achieved by implementing processes using business process technologies such as workflow systems, robotic process automation, or blockchain technologies~\cite{vom2021business}. Often, such process automation results in drastic improvements in process performance indicators. For instance, major companies report cost savings of 50\% thanks to the implementation of robotic process automation~\cite{marek2021process} and 3,000 saved person-hours per month~\cite{ludacka2021digital}.

So far, most BPM research has been concerned with getting process automation done. While the challenges of doing process automation deserve attention, it is equally important to investigate which short and long-term effects process automation can entail. For instance, robotic process automation seems to be a technology with short-term benefits and long-term problems. Maintenance appears to be difficult due to complicated governance and loss of knowledge~\cite{eulerich2022dark,noppen2020keep}. Little of these issues are appropriately reflected in BPM research. A theory explaining process automation's different effects is missing so far.

In this paper, we address this research problem from a theoretical angle. Our approach begins with a review of automation effects observed in prior research on human-automation interaction. However, since most of this research focuses on human control of complex technical systems, its findings cannot be directly applied to process automation. Therefore, we examine the extent to which effects described in research on human-automation interaction are relevant to process automation and derive research areas for better understanding in the future.

This paper is structured as follows: Section 2 provides background information on our research, including an overview of process automation and a discussion of the general perspectives that have developed in research on human-automation interaction. Section 3 outlines our methodological considerations for reviewing the literature. Section 4 presents our findings based on the review of the literature. Section 5 discusses how our findings can inform research on business process automation. Finally, Section 6 concludes the paper with an outlook on future research.

%% file: sections/background.tex
In this section, we discuss what process automation is and which general perspectives on automation have been developed in the human factors literature. This discussion serves as a foundation for our subsequent literature review and identification of process automation's effects.

\subsection{Business Processes and Process Automation}
\begin{figure}[hbt]
    \centering
    \includegraphics[width=0.8\textwidth]{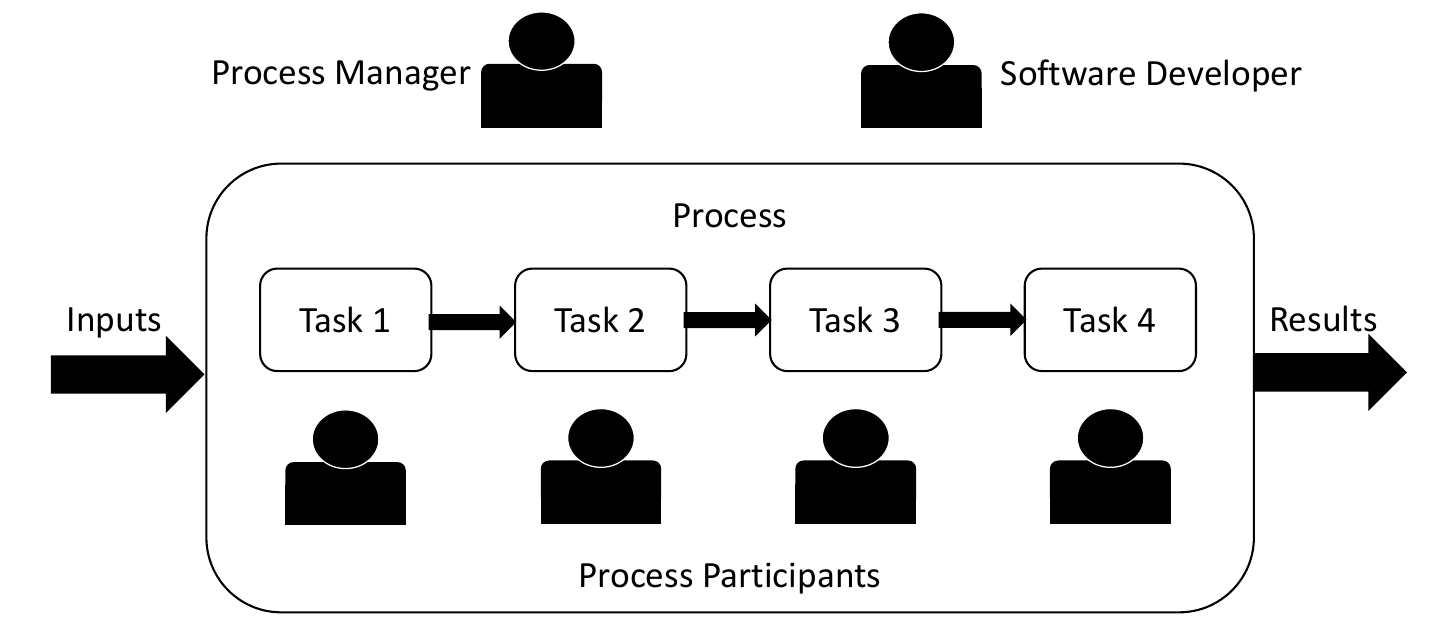}
    \caption{Process, Tasks, Process Participants, Software Developer, and Process Manager}
    \label{fig:process}
\end{figure}

First, we describe how business processes and process automation are related. To this end, we refer to Figure~\ref{fig:process}, a simplification of~\cite[Ch.1]{dumas2018fundamentals}. A business process receives inputs and is targeted toward achieving some desired results. A business process is typically decomposed into tasks for which different process participants are responsible. The performance of the process is monitored by a process manager, who is also responsible for initiating redesigns of the process if desired results are not achieved. The process manager can commission projects towards the automation of the process, which software developers implement.

Automation refers to the machine execution of tasks that humans do not wish to perform or cannot perform as accurately or reliably as machines~\cite{parasuraman2000model}. In line with this statement, we define \emph{process automation} as assigning at least one task or at least one control flow link between two tasks to a machine. In this paper, we focus on information systems as a specific class of machine. Control flow is often automated using enterprise systems, workflow systems, or robotic process automation systems~\cite[Ch.9]{dumas2018fundamentals}.
Information systems that perform tasks previously performed by humans can range from simple tools, such as calculators, to complex systems, such as artificial intelligence and machine learning. 

\subsection{General Perspectives on Automation}
Research on human-automation interaction has developed several useful concepts for investigating automation. 
A key observation of this research discipline is that automation comes in various forms and sizes and thus requires a conceptual differentiation to understand various effects on human performances. Parasuraman's process automation model distinguishes four types of tasks and corresponding technology to substitute human performance: inquiring and presenting information, processing and analyzing information, decision and action selection, and (physical) action implementation~\cite{parasuraman2000designing}. 
\begin{description}
\item[Information presentation automation] includes  generating dashboards or reports to present real-time insights into operational performance or sales figures to users.
\item[Information processing automation] involves calculation, data analysis, automatic reasoning, or natural language processing of large data volumes to identify patterns that are difficult to detect manually. 
\item[Decision automation] can involve using expert systems or decision support tools to provide recommendations based on data analysis and decision rules to assist humans in making complex decisions.
\item[Physical automation] includes assembly line robots in manufacturing plants to improve production efficiency and quality or the use of automated vehicles in logistics to transport goods without human intervention.
\end{description}
More complex systems can provide several of these automation functions, just like autonomous cars require to assess information from the vehicle surroundings, analyze these to decide on how to drive properly, and then actually physically enact them.   

In addition to the types of tasks, Parasuraman et al. proposed a model that includes ten levels of human interaction with automation~\cite{parasuraman2000model}. The levels range from no automation (Level 0) to full automation (Level 9), with Level 10 being a hypothetical level that involves no human interaction. While higher levels of automation reduce the individually perceived workload, this comes at a risk of less situational awareness and a higher chance of missing system failures~\cite{onnasch2014human,wickens2010stages}. 
Overall, each form of automation has the potential to increase efficiency, reduce errors, and enhance safety in various industries. However, it is important to consider the social, psychological, and economic implications of automation, particularly how it affects human labor and job security.

Research on human-automation interaction has discussed automation in various settings. We observe, though, that tasks reflected in this research discipline typically focus on sensor-motoric tasks, often under emergency conditions. Several of the categories established by this literature have face value in the light of process automation, for instance, skill decay. The question begs, however, to which extent research insights on a.o. nuclear power plant disasters and plane crashes can be transferred to an office work area largely supported by information systems and corresponding process automation. We review and structure the human automation interaction literature to address this question and discuss its applicability to process automation.

%% file: sections/methodology.tex
This section discusses our methodology for conducting a representative literature review to collect the main effects of automation \cite{brocke2009reconstructing,cooper1988organizing}. This review aims to identify the effects of process automation on humans involved in business processes. To this end, we turn to engineering psychology, a research field investigating automation in general and diverse settings~\cite{wickens2015engineering}. We deliberately focus on articles published in its flagship journal \emph{Human Factors}, published since 1958, for the paper selection of our literature review. This focus is motivated by the fact that this journal published the most seminal works of engineering psychology. We identify \textit{human-automation interaction} as the key term for our search. Its relevance emerges from process automation applying automation technology for repetitive tasks and coordination. Often, this yields a partial or semi-automated solution. As search string, we therefore use ``\emph{human-automation interaction}'' as well as the combination of ``\emph {human interaction}'' and ``\emph{automation}''.

We were aware that the term \emph{automation} in \emph{Human Factors} is broader than \emph{process automation}. The former includes, for example, areas such as autonomous driving or flight simulation. We believe that the interaction of humans and automation and the effects on humans in these areas need to be investigated if and to what extent they are also relevant to process automation. 

The search was conducted in February 2023 and yielded 75 papers. To be included, a paper had to focus on phenomena that could be related to process automation and describe the interaction between humans and technology. We excluded papers describing specific technical mechanisms, solely physical interactions, or focused on highly specialized industries. For example, sensomotirical aspects such as pressing the right button were excluded because they are less relevant for workflows or robotic process automation. For similar reasons, we excluded papers  focusing on highly specific domains, such as space science.  After selection, 52 papers are included in the analysis.



The first author read all of these papers in their entirety and collected the phenomena and effects of human-automation interaction. To structure the effects, a data-driven qualitative inductive content analysis approach was followed~\cite{gibbs2007thematic}. In several iterations with the author team, all phenomena from the literature were mapped to the basic principles Prerequisites, Phenomena, Consequences, known from the grounded theory approach~\cite{corbin1990grounded}. Since our main object of investigation is human automation, we further distinguish between human-driven aspects and technology-driven aspects within these three categories. In the following, we explain the overall model with detailed examples from the literature.

%% file: sections/findings.tex

The literature on human-technology interaction describes phenomena categorized into interaction preconditions, main interaction phenomena, and resulting consequences (see Fig. 2). Successful interaction requires both humans and technology to meet certain basic requirements. The human must possess knowledge of and trust in the technology, while the technology must fulfill basic functionalities and design principles that cater to human specifics.

During the interaction, effects are observed on both the human and technical sides. These effects are influenced by preconditions and are inherent to the interaction and communication between the two parties. For instance, humans tend to blame technology for failure, particularly when performance expectations are unrealistic. The transparency and communication of the technology affect human reactions.

Long-term effects are visible on both the human and technical side, influenced by previous interactions and preconditions. Successful process automation can lead to humans relying on the system and losing process knowledge due to inactivity. However, it can be challenging for humans to react adequately in cases of technology failure.

\begin{figure}[!h]
    \centering
    \includegraphics[width=1\textwidth]{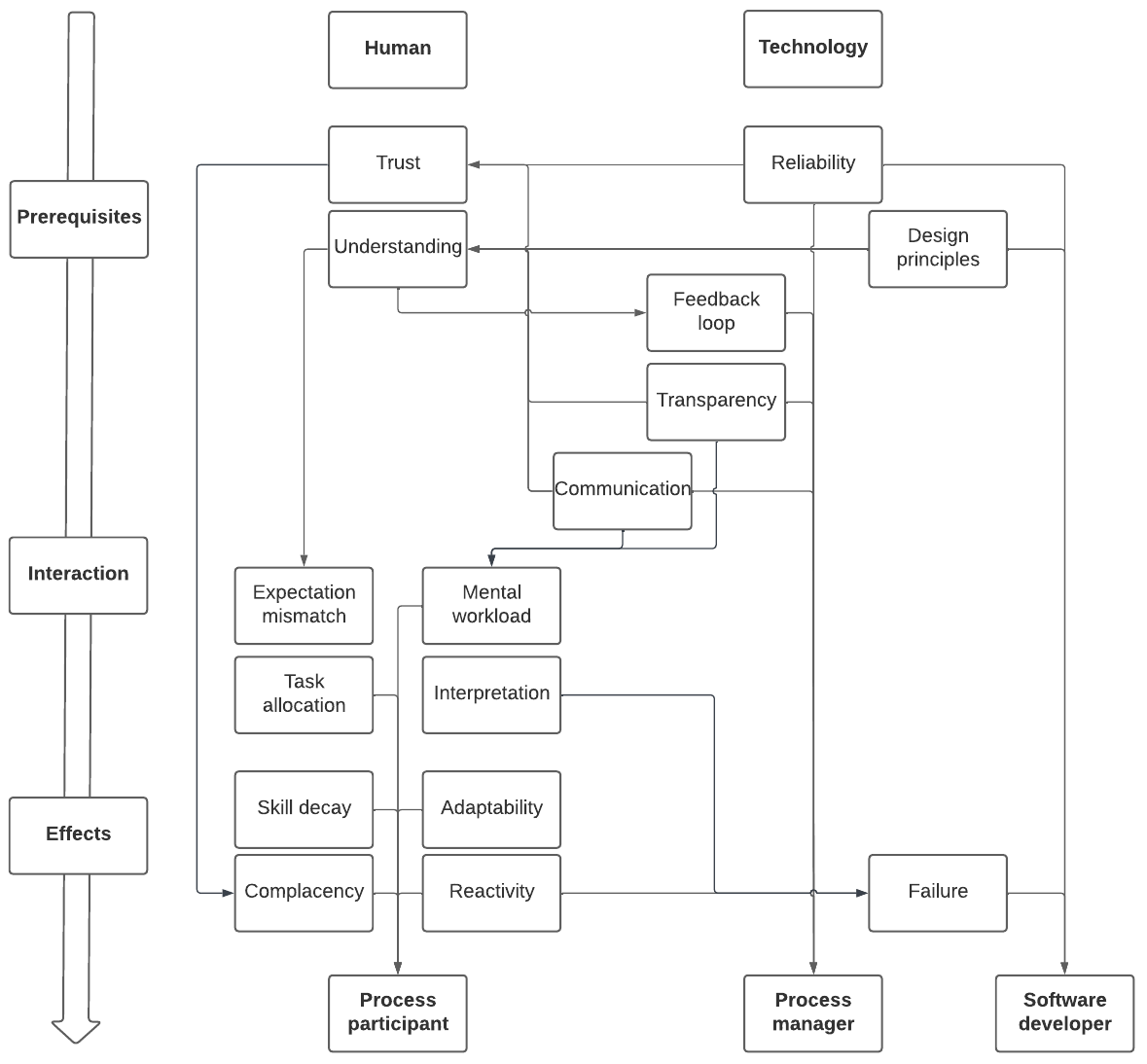}
    \caption{Overview of all effects found in human-automation interaction literature}
    \label{fig:overview}
\end{figure}

In the following, we present the three levels of automation interaction in detail (see Fig. \ref{fig:overview}) from a human and technical perspective. Note that our list is not exhaustive as we focus on the phenomena discussed in the Human Factors literature. We identified the relationships in a unidirectional manner focusing on the influence of an aspect on another. An example is that the \textit{understanding} of an automation influences the \textit{expectation mismatch} of a human.

\subsection{Prerequisites}
The first level, prerequisites,  pertains to the fundamental steps to facilitate a successful interaction between humans and automation technology. It emphasizes the human prerequisites for interaction, technological design principles, and essential properties. Our analysis has identified \textit{understanding} and \textit{trust} as crucial human requirements, while \textit{transparency}, \textit{reliability}, \textit{design principles}, and the \textit{feedback loop} are significant technological considerations.

\mypar{Humans need a certain understanding of technology}
To effectively work with automation technology, humans benefit from \emph{understanding} the underlying technological mechanisms. It has been found that many humans are generally under-informed in that regard~\cite{skraaning2021human}. Due to the technological advancements in automation, automation technology has become more independent, and humans are no longer as 'in the loop' as before. These advancements can increase the complexity for humans, which can result in gaps and misconceptions. Consequently, automation technology often surprises humans as it behaves in a manner that humans neither anticipate nor understand~\cite{victor2018automation}. 
Therefore, understanding automation technology is a key concern and refers to the in-depth knowledge a person has about how it works, what it can do, and its limitations. Understanding the technology is considered more important than its reliability and competence in real-world scenarios~\cite{balfe2018understanding}. 

\mypar{Humans need a certain trust in technology}
Another aspect that needs to be accounted for is the \emph{trust} humans have in automation technology. The level of trust determines to what extent humans are willing to accept automation technology and how well they interact with it in interactive settings~\cite{kaplan2021trust}. 
A study on the usage of artificial intelligence devices has shown that the trust of humans heavily depends on the technology's transparency, reliability, and compatibility~\cite{ismatullaev2022review}. 
Humans tend to trust automation technology as they perceive it as an instrument with superior analytical capabilities and can outperform humans~\cite{parasuraman2010complacency}. However, in scenarios where expert knowledge is required, the human tends to trust the human expert more. This highlights that human perception can vary based on the complexity of the automation scenario and that establishing trust is particularly important in these scenarios~\cite{madhavan2007effects}.

\mypar{Technology needs to be transparent and reliable}
For an effective interaction between humans and automation technology, both \emph{transparency} and \textit{reliability} are important concerns. Transparency may relate to various features of automation technology, including its capabilities, responsibilities, activities, and goals. Depending on the specific automation scenario, transparency of different features might be important. For example, a study on automation in a nuclear power plant has illustrated that a high level of transparency regarding which automation components are currently running, how they work, and how they interact with other components has led to an increased supervision performance of the operators~\cite{skraaning2021human}.
Besides transparency, also \emph{reliability} plays an important role. In general, automation technology is considered reliable if it performs consistently and accurately from the user's point of view. Reliability has been found to significantly influence people's perceived trustworthiness of technologies~\cite{lyons2021trusting}.

\mypar{Technology needs to follow human-centric design principles}
Another aspect from the technology perspective is the automation design itself.
It is generally difficult to design automation technology to make it clear and understandable to humans~\cite{skraaning2021human}. The implications of ``bad'' design can be severe. In some cases, the effort for using automation technology can outweigh its advantages (in terms of time savings, etc.), meaning that not using the automation technology is perceived as the better option~\cite{eriksson2017driving}. In other cases, a non-suitable design might lead to human errors, which, at least in critical domains such as aviation, should be prevented at all costs~\cite{marquez2017measuring}.   

Against this background, it is important to follow the human-centric \emph{design principles}, which provide recommendations for human-automation interaction~\cite{jamieson2005designing} as well as the user experience~\cite{heymann2007formal}. Among others, relevant aspects include the format of the user interface~\cite{bartlett2019no}, the position of text and images~\cite{sylla1988human}, and the amount of data that is presented to humans~\cite{gillan1998guidelines}.  
Studies have shown that providing the right visual information to humans can increase their overall performance~\cite{mcguirl2006supporting}. 




\mypar{Technology has to provide an open framework for human input} The \emph{feedback loop} of automation technology, i.e., the possibility to include and consider human inputs, plays an important role in the overall performance of human-automation interaction. In this context, \emph{adaptive automation} can incorporate the human response in the automation design~\cite{feigh2012toward}. By including human aspects, such as frustration, satisfaction, motivation, or confidence, to improve automation design, a higher performance, better results, and lower workload for the human can be achieved~\cite{prinzel2003effects,yang2018affect}. 

For complex socio-technical environments, the framework \emph{ecological interface design} captures additional aspects, such as social and physical characteristics as well as the natural relationship in human-automation interaction~\cite{vicente2002ecological}. 
This approach aims to enhance usability and safety and has been examined in the context of semantic mapping~\cite{vernon2002ecological}. 

Another framework that should be considered when designing human-focused automation is \emph{cognitive engineering} as it aligns the automation design within the cognitive abilities and constraints of a human~\cite{woods1988cognitive}. This has been used in work domain analysis~\cite{naikar2001evaluating} and in a systemic model of computer response~\cite{whang2003preparing} to understand how humans process information, make their decisions and perform their tasks and use this as input to design an effective and human-oriented design. 

\subsection{Interaction}
This section focuses on the various facets of human-automation interaction. It elucidates human behavior during the interaction and the anticipated response of the technology. Our analysis has identified expectation mismatch, interpretation, blame, acceptance, task allocation, and mental workload as the primary human aspects, while communication and justification are significant technological considerations.

\mypar{Humans tend to perceive an automation technology incompletely}
The way humans perceive a given automation solution, e.g., in terms of its usefulness, is based on their knowledge of the underlying technology. This can result in an \emph{expectation mismatch}~\cite{madhavan2007effects}, meaning that humans may have too high expectations concerning the automation technology's capabilities. This, in turn, may lead to a lower human-automation interaction performance, e.g., because the human is not expecting the need to deal with automation failure~\cite{victor2018automation}. In these situations, humans tend to blame the automation technology and do not recognize their own responsibility as they perceive the automation technology as superior in terms of capabilities~\cite{douer2022judging}. This effect is even stronger when the degree of automation is higher~\cite{furlough2021attributing}.   



\mypar{Humans tend not to interpret information thoroughly}
Many automation solutions present data to humans. However, humans tend not to thoroughly examine the available data and do not always interpret it correctly~\cite{parasuraman2010complacency}. This can lead to the human interacting with automation technology in an unintentional manner, resulting in unintended and incorrect actions by the system~\cite{sarter1995world}. 

 

\mypar{Humans tend to manage their task inefficiently}
When interacting with automation technology, humans need to focus on their \emph{task allocation}, i.e., the tasks that have been assigned to them. A laboratory experiment has demonstrated that whenever automation greatly exceeds human capabilities, humans feel that they contribute too little to the overall task. As a result, they may intervene more often than required~\cite{douer2022judging}. Especially in multi-tasking scenarios, this has been found to reduce human performance~\cite{parasuraman2010complacency}. 

Another concern related to human-automation interaction is transitioning from automation to human, i.e., takeovers. Here, it has been found that the (perceived) absence of time pressure may lead to longer transition times. Also, multi-tasking leads to a slower response to resume control~\cite{eriksson2017takeover}. In this context, human \emph{mental workload} plays an important role as it determines human cognitive abilities and resources to successfully perform a task. 
Studies have shown that overall interaction with automation technology can initially result in an extensive mental workload for a human compared to not involving automation at all~\cite{collet2003assessing}. This phenomenon, however, decreases over time through a learning process. A way to reduce the mental workload for humans is to transparently present relevant information for the employed automation solution~\cite{van2022agent}. 


\mypar{Technology should communicate with humans}
Effective collaboration between humans and automation technology requires a certain level of \textit{communication} from automation technology to humans. This allows humans to develop a high level of trust and understanding, which results in the human being less content to monitor and intervene with an automation unnecessarily~\cite{balfe2018understanding}. One example is the proper communication of system state uncertainties, as these have a direct effect on the human mental workload, visual attention, and situational awareness~\cite{johnson2017closed}. If a human knows that critical situations will be communicated properly, they can focus on their other tasks without spending time monitoring detailed parameters. Another example relates to communicating social intent~\cite{lyons2021trusting}. If humans know that automation technology considers human well-being explicitly (e.g., by preventing accidents in a production context), this increases the human's trustworthiness.

\subsection{Effects}
This section details the various effects of human-automation interaction. It describes how humans are influenced during and by the interaction, as well as the implications for the technology itself. Our analysis has identified reactivity, adaptability, and skill decay as the primary human aspects, while failure is a significant technological consideration.

\mypar{Humans are affected in their readiness to intervene}
As automation technology advances and becomes more robust, humans tend to become less aware of their situation and are less likely to take over manual control when needed~\cite{endsley2017here}. This phenomenon is typically referred to as \emph{skill decay}. Skill decay can lead to \emph{vicious cycles} because humans lose their skills to take over in the course of time. If they, however, need to take over (because the automation technology fails), they might not have the ability to do it, leading to an even higher level of dependency on the automation technology~\cite{lee2008review}. This means that the ability of humans to intervene in critical processes must be both established and maintained~\cite{wickens2015using}.      


\mypar{Humans are affected by automation complacency}
An additional effect that occurs with an increasing level of automation is \emph{automation complacency}. This phenomenon refers to a situation where humans become too comfortable and complacent with an automation technology~\cite{parasuraman2010complacency}. The consequence of automation complacency is the general human expectation that automation technology will work, without knowing or understanding whether this will be the case. A related effect is \emph{automation bias}, which arises when humans blindly rely on automation technology without actively monitoring and validating its activities. Both effects originate in the human over-trusting automation technology and may pose severe risks for the performance of human-automation interaction~\cite{parasuraman2010complacency}. A way to manage complacency is complacency modeling, which can help predict the effects of different types of imperfect automation technology~\cite{wickens2015using}.

\mypar{Humans are affected by automation changes}
Over time, the nature of human-automation interaction may change due to technological advancements. Therefore, an effective human-automation performance requires the human to \textit{adapt} to these changes. In many cases, this is a question of additional training~\cite{lee2008review}. In some cases, however, humans have also been able to adapt without additional training. For example, a study with helicopter pilots has shown an increased human performance when the pilots were presented with additional information~\cite{innes2021effects}. 


\mypar{Technology may fail}
The interaction between humans and automation technology may cause \emph{automation failure}. In such a case, the question is whether the human or the automation solution should react to the failure task. A study examining the probability of missed failures (false negatives) and false alarms (false positives) showed that, for time-critical scenarios, an automation technology might fit better to handle the failure, whereas in most other cases a human should take over the automation failure~\cite{sheridan2000human}. The \emph{lumberjack analogy} points out that as the level of automation increases, the performance of routine tasks improves, but the monitoring and reactivity of the human to failure scenarios significantly decreases~\cite{sebok2017implementing}. 

%% file: sections/discussion.tex
Our review of human factors research clarifies the complex interaction between humans and automation technology. 
We identified three focal areas: interaction prerequisites, main interaction phenomena, and interaction effects. 
The specific aspects of these focal areas have implications for human-automation interaction in the context of business process automation. The goal of this section is to make these links explicit and highlight how our findings can inform research on business process automation. To structure our discussion, we use the three roles related to process automation, namely process participants, process managers, and software developers. 

\subsection{Process Participant}
The prerequisites of human-automation interaction are \emph{understanding} and \emph{trust}. We argue that these are equally applicable to process participants in the context of process automation. A process participant benefits from having foundational knowledge about the automation solution in semi- and fully-automated scenarios to effectively work with it. During the interaction itself, 
the process participant can have an \emph{incomplete understanding} and may \emph{misinterpret information} regarding an automation solution. This aspect is accelerated because the division of labor hinders process participants in their understanding of the whole process~\cite{dumas2018fundamentals}, even if no automation is in place. These aspects can be addressed through training and education initiatives that teach the process participant to work efficiently with automation technology. In addition, the effects on \emph{mental workload} and \emph{task allocation} are worsened, given that the process participant might take over tasks in multiple processes. 
The exposure to \emph{automation change} is as relevant for the process participant. The process changes over time, defining new requirements on the technology, and corresponding changes likely impact participants and their role in the process. 
Therefore, relevant research questions from the perspective of the process participants include:

\setdescription{itemsep=6pt}
\begin{description}

    \item[PP1:] What are the gaps in foundational knowledge with respect to automation solutions? To address the problem of insufficient foundational knowledge, it is important to identify which aspects process participants typically struggle with and which of these aspects may lead to lower performance.  
        
    \item[PP2:] How can effective training and education initiatives be developed to support process participants who work with automation technologies? A key concern in this context will be mechanisms to develop understanding and trust, with a specific focus on process change. 
    
    \item[PP3:] What is the impact of automation change on the role and responsibilities of process participants? Change in this context may have a variety of implications. Among others, it might be necessary to reestablish trust as well as the human understanding of the automation solution. It might be interesting to also connect these aspects to typical challenges of change management, such as resistance to change.  
    
    \item[PP4:] Which factors lead to incomplete understanding and misinterpreting automation information on process performance? It is important to understand which factors may cause these issues to effectively prevent them. Possible causes may relate to human understanding, the automation design, but also cognitive factors of the human, such as mental workload.   
    
    \item[PP5:] What is the impact of task allocation on mental workload, and what are ways to optimize it in the context of process automation?  The key concern in this context is not to overwhelm humans. Especially when automation technology is introduced or changed, there is a need for careful consideration of the human mental workload, such that the benefits of automation technology are not outweighed by humans struggling with an effective human-automation interaction.  
    
\end{description}



\subsection{Process Manager}

From the standpoint of the process manager, the aspects of \emph{transparency} and \emph{reliability} are just as important in process automation. As automation technology should work as designed, the process manager must ensure it works reliably and communicates its results transparently. In this context, the interplay with technology requires a \emph{human-centric design} that meets the profile and characteristics of the people involved. Additionally, as the technology in process automation might also fail, the process manager needs to manage these failure scenarios with an error-handling mechanism. This includes whether error handling should be performed by a human, such as a process manager or process participant, or by a technology that monitors automation and performs a task if a dedicated condition is encountered.
Therefore, research questions from the perspective of the process manager should include:
\begin{description}

    \item[PM1:] How can process managers select a human-centric design incorporating human trademarks and needs? It is important that the process manager chooses the most suitable technology that fits the needs of their users and can work efficiently in different process contexts.   
    
    \item[PM2:] How can process managers establish a bi-directional communication channel for efficient collaboration and feedback? While the feedback loop has been emphasized as an important aspect, how can process managers establish a communication channel that allows humans to provide feedback to an automation solution and the other way around? 
    
    \item[PM3:] How can process managers handle and manage failure scenarios? Automation failure is (almost) inevitable. Therefore, it is relevant to identify and classify typical failure scenarios. So-called black swan events, i.e., rare but severe errors, might deserve particular attention in this context. 
    
    \item[PM4:] How can process managers measure the effectiveness of different error-handling mechanisms and choose the optimal approach for different types of process automation? Once error-handling mechanisms have been selected (see PM3), it is critical to understand their effectiveness. Again, the criticality of the domain might be an important contextual factor that needs to be considered. 
    
    \item[PM5:] What is the impact of automation on the process manager's role and responsibilities, and what are ways to optimize their performance in the context of process automation? Many of the questions above ultimately focus on the interaction between the user and automation technology. However, it is also relevant to understand how the roles and responsibilities of the process manager are evolving. Among others, process managers might need to monitor how \textit{drift} affects the performance of automation technology or establish governance mechanisms for the different automation solutions employed.    
    
\end{description}

\subsection{Software Developers}
From the perspective of a software developer, the goal is to design automation technology that meets basic design principles and is reliable, robust, and free of failures. The automation technology should provide the necessary features from the human point of view and enable individualization of the interaction. Further research from the perspective of the software developer should include the following research questions:
\begin{description}
    \item[SD1:] How can software developers implement basic design principles to ensure the usability and effectiveness of automation technology? The importance of design principles is well known, but it is necessary to understand how existing design principles affect the usability and effectiveness of automation technology. 
    
    \item[SD2:] How can software developers integrate human feedback into their development process to continuously improve it? Automation technology must evolve over time, e.g., due to drift, technological advances, or changing requirements. Finding ways to effectively integrate user feedback is necessary to continuously improve the automation solutions used.
    \item[SD3:] How must software developers design automation technologies to communicate with humans efficiently and as free of interpretation as possible? Efficient communication positively affects user perception, acceptance, and comprehensibility. 
        
    \item[SD4:] What are mechanisms to implement error handling to manage failure scenarios in process automation? Effective approaches for error handling should be developed, whether performed by users or the automation technology itself. By building on the identified failure scenarios (see PM3), a link between failure scenarios and the error-handling mechanism can be established.  
    
    \item[SD5:] How can software developers choose the optimal balance between automation and human involvement for different types of processes? One of the key questions with ongoing technological advancements is how much humans still need to be involved and whether the benefits from the shift towards automation technology outweigh the cost of its implementation. A comprehensive investigation of this question for various task types and processes is imperative for an outcome-driven usage of automation technology. 
    
\end{description}    

\noindent The identified research directions require an empirical research agenda. Research on human-automation interaction has developed specific research strategies, such as experimental designs, simulation, observational studies, and case studies~\cite{wickens2015engineering}, which can be adopted and reused to this end. Corresponding results will provide the foundation for a theory on process automation.



%% file: sections/conclusion.tex
This paper focused on the human aspects of process automation, recognizing that humans are critical to the success of all process automation scenarios, whether as process participants in semi-automated processes or as process managers overlooking interaction between users and automation technology. While existing research on process automation elaborates on the technological options for implementation, such as robotic process automation or workflow capabilities, we shifted the focus towards the human as a crucial factor for process execution success. To this end, we examined the journal \emph{Human Factors} to identify relevant aspects and effects of human-automation interaction applicable to the field of process automation. 

Our analysis highlighted multiple aspects of both humans and technology, which we classified into the categories prerequisites, interaction, and effects. These aspects illustrate the complexity of the relationship between technology and humans and the diverse factors that can influence their performance. We concluded that it is essential to incorporate human characteristics and trademarks into automation design, establish efficient means of communication between humans and technology, and carefully evaluate the appropriate degree of automation based on its impact on overall process performance and its ability to support human decision-making based on their cognitive abilities.